\documentclass[twocolumn,aps,prc,showpacs,preprintnumbers,amsmath,showkeys,amssymb,superscriptaddress]{revtex4}

\usepackage{graphicx}
\usepackage{dcolumn}
\usepackage{bm}

\newcommand{\nuc}[2]{\ensuremath{^{\text{#1}}\text{#2}}}

\begin{document}

\preprint{APS/123-QED}

\title{Neutron knockout of \nuc{12}{Be} populating neutron-unbound states in \nuc{11}{Be}}

\author{W.~A.~Peters}
 \email{wapeters@nuclearemail.org}
   \affiliation{Department of Physics and Astronomy, Rutgers, \\ State University of New Jersey, Piscataway, NJ 08854, USA.}
  \affiliation{Department of Physics \& Astronomy, Michigan State University,
               East Lansing, MI 48824, USA}
  \affiliation{National Superconducting Cyclotron Laboratory,
               Michigan State University, East Lansing, MI 48824, USA}

\author{T.~Baumann}
  \affiliation{National Superconducting Cyclotron Laboratory,
               Michigan State University, East Lansing, MI 48824, USA}
\author{B.~A.~Brown}
  \affiliation{Department of Physics \& Astronomy, Michigan State University,
               East Lansing, MI 48824, USA}
  \affiliation{National Superconducting Cyclotron Laboratory,
               Michigan State University, East Lansing, MI 48824, USA}

\author{J.~Brown}
  \affiliation{Department of Physics, Wabash College, Crawfordsville, IN 47933, USA}

\author{P.~A.~DeYoung}
  \affiliation{Department of Physics, Hope College, Holland, MI 49423, USA}

\author{J.~E.~Finck}
  \affiliation{Department of Physics, Central Michigan University, Mt. Pleasant, MI 48859, USA}

\author{N.~Frank}
  \altaffiliation[Currently at ]{Department of Physics,
Augustana College, Rock Island, IL 61201, USA}
  \affiliation{Department of Physics \& Astronomy, Michigan State University,
               East Lansing, MI 48824, USA}
  \affiliation{National Superconducting Cyclotron Laboratory,
               Michigan State University, East Lansing, MI 48824, USA}

\author{K.~L.~Jones}  
    \altaffiliation[Currently at ]{Department of Physics \& Astronomy, University of Tennessee, Knoxville, TN 37996, USA}
   \affiliation{Department of Physics and Astronomy, Rutgers, \\ State University of New Jersey, Piscataway, NJ 08854, USA.}

\author{J.-L.~Lecouey}
  \altaffiliation[Currently at ]{Laboratoire de Physique Corpusculaire, ENSICAEN IN2P3, 14050, Caen, Cedex, France}
  \affiliation{National Superconducting Cyclotron Laboratory,
               Michigan State University, East Lansing, MI 48824, USA}

\author{B.~Luther}
  \affiliation{Department of Physics, Concordia College, Moorhead, MN 56562, USA}

\author{G.~F.~Peaslee}
  \affiliation{Department of Physics, Hope College, Holland, MI 49423, USA}

\author{W.~F.~Rogers}
  \affiliation{Department of Physics, Westmont College, Santa Barbara, CA 93108, USA}

\author{A.~Schiller}
  \altaffiliation[Currently at ]{Department of Physics \& Astronomy, Ohio University, Athens, OH 45701, USA}
  \affiliation{National Superconducting Cyclotron Laboratory,
               Michigan State University, East Lansing, MI 48824, USA}

\author{M.~Thoennessen}%
  \affiliation{Department of Physics \& Astronomy, Michigan State University,
               East Lansing, MI 48824, USA}
  \affiliation{National Superconducting Cyclotron Laboratory,
               Michigan State University, East Lansing, MI 48824, USA}

\author{J.~A.~Tostevin}
  \affiliation{Department of Physics, Faculty of Engineering and Physical Sciences, University of Surrey, Guildford, Surrey GU27XH, U.K.}

\author{K.~Yoneda}
  \altaffiliation[Currently at ]{RIKEN Nishina Center, Wako, Saitama 351-0198, Japan}
  \affiliation{National Superconducting Cyclotron Laboratory,
               Michigan State University, East Lansing, MI 48824, USA}


\date{\today}

\begin{abstract}
Neutron-unbound resonant states of \nuc{11}{Be} were populated in neutron knock-out reactions from \nuc{12}{Be} and identified by \nuc{10}{Be}--n coincidence measurements. A resonance in the decay-energy spectrum at 80(2)~keV was attributed to a highly excited unbound state in \nuc{11}{Be} at 3.949(2)~MeV decaying to the $2^+$ excited state in \nuc{10}{Be}. A knockout cross section of 15(3)~mb was inferred for this 3.949(2)~MeV state suggesting a spectroscopic factor near unity for this $0p 3/2^{-}$ level, consistent with the detailed shell model calculations.

\end{abstract}


\maketitle

Several recent experiments have mapped the level structure of \nuc{11}{Be}. Hirayama {\it et al.} \cite{hira05} observed the $\beta$-delayed 
neutron decay from polarized \nuc{11}{Li}, identifying neutron-unbound levels in \nuc{11}{Be} and assigning spin and parity to each. Previous neutron knockout experiments have identified additional levels, and highlighted significant mixing with $sd$-shell states \cite{pain06prl,navi00prl}. We also report on neutron-unbound excited states in \nuc{11}{Be} populated by neutron knockout from \nuc{12}{Be} and investigated by in-beam neutron-decay spectroscopy. These data show a resonance at a decay energy of 80(2)~keV indicating population of the known $3/2^-$ state at 3.949(2)~MeV in \nuc{11}{Be} decaying to the first $2^+$ state in \nuc{10}{Be} via neutron emission. The uncertainty of the measured energy for this state is significantly improved over the previous accepted value \cite{ajze90}. The measured knock-out cross section of 15(3)~mb implies a spectroscopic factor near unity for this $3/2^-$ state.
\par
\begin{figure}[t]
\centering
\includegraphics[width=3.0in]{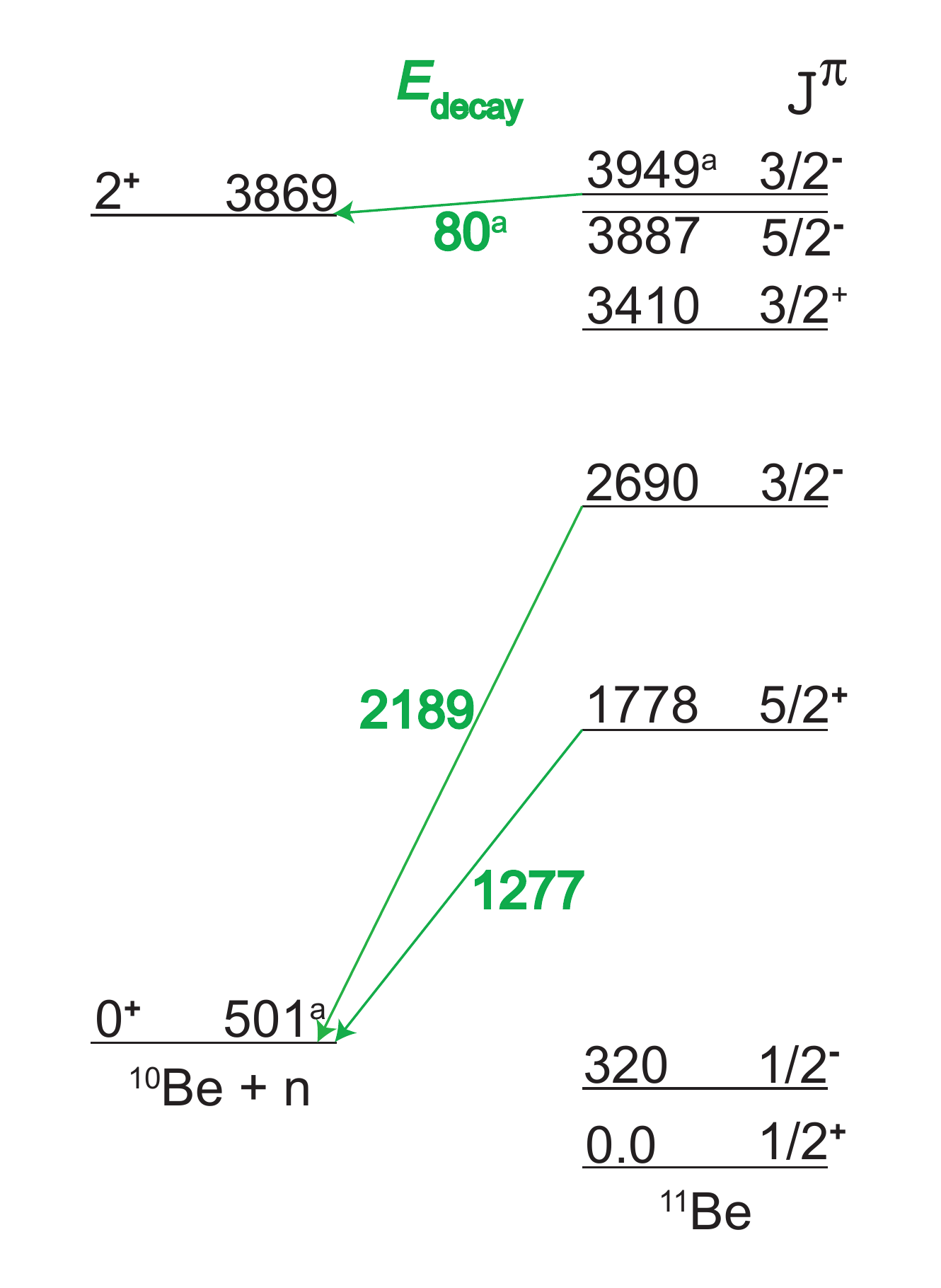}
\caption{(Color online) \nuc{11}{Be} level scheme up to 4~MeV including the first two states in \nuc{10}{Be}. The neutron decay energies observed in this experiment to the \nuc{10}{Be} ground state or first excited state are labeled. Energies are given in keV along with known spin and parity assignments. ($^{a}$ denotes values derived in the current work incorporating the recently remeasured separation energy from Ref. \cite{ring09}).}
\label{12be_shells}
\end{figure}

The reports of Hirayama {\it et al.} \cite{hira05}, Aoi {\it et al.} \cite{aoi97npa}, and Morrissey {\it et al.} \cite{morr97npa} from $\beta$-decay of \nuc{11}{Li}, noted excited states in \nuc{11}{Be} including (1.778~MeV)($J^\pi=5/2^+$), (2.690~MeV)($J^\pi=3/2^-$), and (3.949~MeV)($J^\pi=3/2^-$) that are also observed in this work. Additionally, Navin {\it et al.} \cite{navi00prl} demonstrated the importance of $sd$ intruder states to understanding the structure of \nuc{11}{Be} by using neutron-knockout reactions from \nuc{12}{Be} to populate the $1/2^+$ and the $1/2^-$ states in \nuc{11}{Be}. These levels from $\nu(1s_{1/2})^{2}$ and $\nu(0p_{1/2})^{2}$ valence neutron configurations in \nuc{12}{Be} were found to be populated with nearly equal probability. This significant shell-level mixing with the $sd$-shell, the subsequent fragmentation of simple single-particle strengths \cite{kana05epj,hama07}, $\alpha$-particle clustering, and resulting deformation, contribute to the disappearance of the eight-neutron magic shell gap in \nuc{12}{Be}. Pain {\it et al.} further identified a possible resonance at approximately 3.5~MeV decay energy. They also observe a narrow resonance near zero due to a state (or two states) in \nuc{11}{Be} at about 4~MeV excitation energy that subsequently decay via neutron emission to the first excited $2^+$ state of \nuc{10}{Be} at 3.368~MeV, but these paths could not be well defined by their data because of limitations in their experimental setup. We employed the neutron-knockout technique of References \cite{navi00prl,pain06prl} using the Modular Neutron Array (MoNA) \cite{mona1,mona2}. Figure \ref{12be_shells} displays the level scheme for the low-lying energy levels in \nuc{10}{Be} and \nuc{11}{Be} including the neutron decay energies seen in the present experiment.

\par
The experiment consisted of a primary beam of \nuc{18}{O} accelerated to 120~MeV/nucleon with the Coupled Cyclotron Facility \cite{ccf} at the National Superconducting Cyclotron Laboratory; this beam impinged onto a 1080~${\rm mg/cm^{2}}$ \nuc{9}{Be} production target. The secondary beam of 90~MeV/u \nuc{12}{Be}, produced by fragmentation, was separated with the A1900 fragment separator \cite{a1900} utilizing a 750~${\rm mg/cm^{2}}$ acrylic achromatic wedge degrader installed at the dispersive image. The average intensity of the \nuc{12}{Be} beam was about 60,000 particles per second, with a momentum spread of $\pm 0.5$\% and a purity of over 99\%.

\par
The secondary beam was directed onto a 102~${\rm mg/cm^{2}}$ \nuc{9}{Be} reaction target. Charged particles were deflected by the large gap Sweeper magnet \cite{sweeper1,sweeper2} and the neutrons were detected by MoNA \cite{mona1,mona2}. The setup and the charged-particle detectors after the Sweeper magnet are described in Figure 4 of Ref.~\cite{mona1}. Additionally, a steel blocker was installed in front of the first CRDC to protect it from the low-momentum tail of the unreacted \nuc{12}{Be} beam.

\par
The energies of the neutrons were calculated from the flight time between a timing detector in front of the reaction target and MoNA, located at zero degrees and positioned 8.2~m from the reaction target. Their angles relative to the beam axis were assigned by the first interaction point in MoNA. Timing the arrival of the light at each end of neutron detector bars yields a horizontal position with a standard deviation of 3~cm. The vertical and longitudinal position resolution is 5~cm (one half the bar width and height of 10~cm) \cite{peters_thesis,mona1}.

\par
The directions of the charged particles behind the Sweeper magnet were measured by two Cathode Readout Drift Chambers (CRDCs). The position resolution of the CRDCs was 1.5~mm in the horizontal dispersive plane. The energy and emission angle of each fragment at the reaction target was calculated using a transformation matrix constructed from the measured magnetic field maps of the Sweeper \cite{frank_thesis} using the beam physics code package {\sc cosy infinity} \cite{cosy,cosy2}. The elemental identification of the charged fragments was based on energy loss in a plastic scintillator downstream of the two CRDCs. Isotopic separation of the beryllium nuclei was based on the measured horizontal angle determined by the two CRDCs and the fragment flight time between the timing detector at the target to the d$E$ scintillator as in Ref. \cite{chri08npa}. The results presented below are based on events with a neutron in coincidence with a \nuc{10}{Be} fragment. This coincidence gate yields a clean neutron spectrum with little background. The decay energy can be determined by subtracting the mass of the decay products from the invariant mass of the neutron--fragment system as described in Ref. \cite{schi05prc}. The neutrons are moving near beam velocity (90~MeV in the current experiment) and are forward focused. This results in a neutron acceptance of 60\% for decay energies less than 2.5~MeV. The resolutions described above propagate through the invariant-mass equation and broaden the resolution of the decay energy as the square-root of the energy; from a standard deviation of 75~keV at 300~keV, to 200~keV for a decay energy of 1500~keV \cite{peters_thesis}.

\begin{figure}[t]
\includegraphics[width=3.2in]{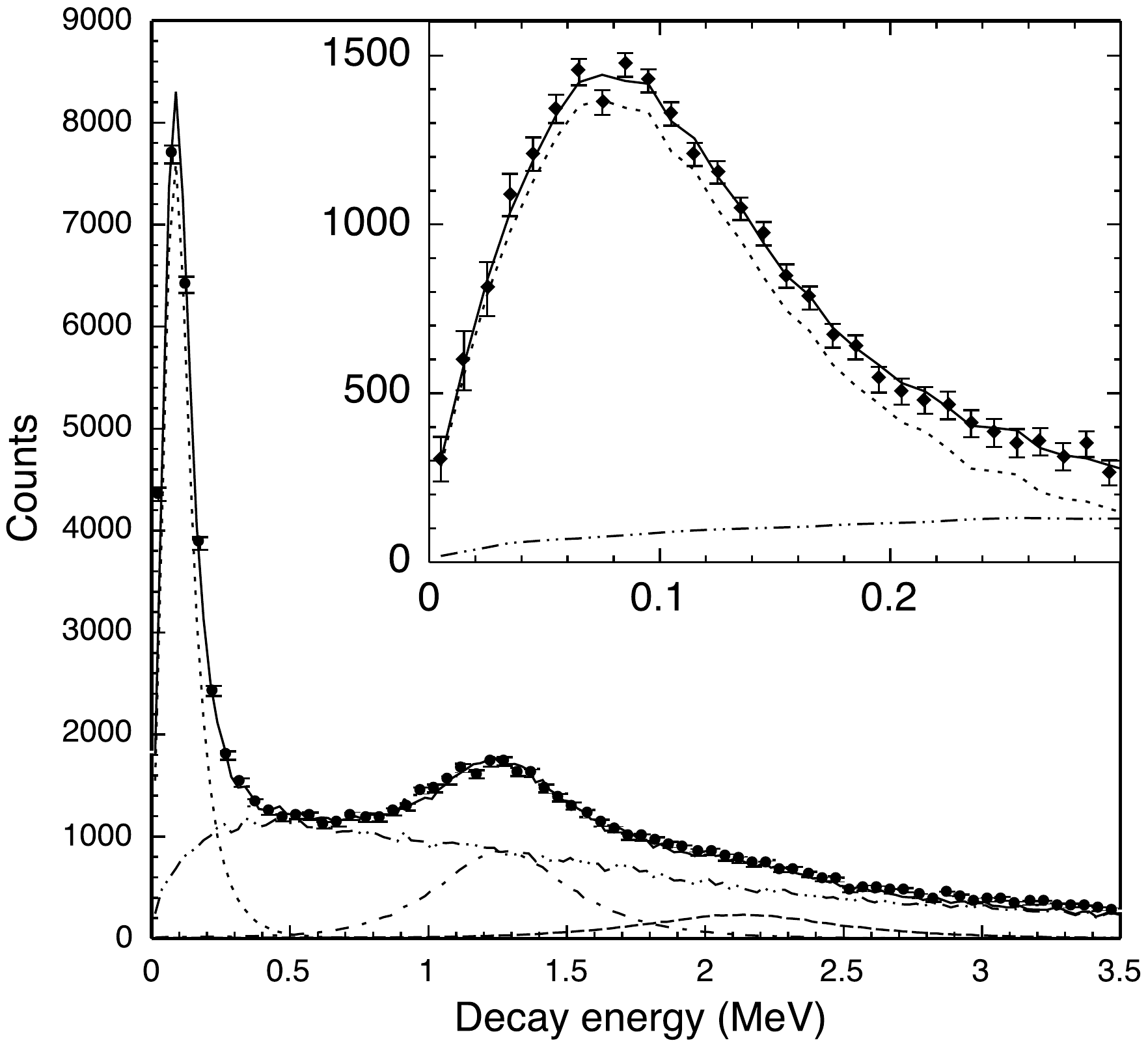}
\centering
\caption[\nuc{12}{Be} experiment fit to data allowing individual amplitudes to adjust.]{Decay energy spectrum from \nuc{10}{Be}--neutron coincidence data. The simulation (solid line) is the sum of three resonances with decay energies of 80~keV (dotted line), 1.28~MeV (dot-dashed line), and 2.19~MeV (dashed line). In addition, a non-resonant background component (double-dot dashed line) was included. The insert shows a separate fit to the low energy range with higher fidelity, confirming only one low energy peak at 80~keV.}
\label{03048a_free_fit}
\end{figure}

The decay energy spectrum is shown in Fig.~\ref{03048a_free_fit} and two prominent peaks are indicated, one produced by a low-energy decay (less than 100~keV), and the other with an energy of 1.28~MeV. The overall shape of the spectrum is similar to the decay energy spectrum presented in Ref.~\cite{pain06prl}. A detailed simulation of the data, as described below, further indicates the presence of a broad resonance with decay energy of 2.19~MeV.

\par
Monte Carlo simulations were performed which incorporated the geometric acceptances and measured resolutions of the neutron and charged particle detectors. The resonances were modeled by Breit-Wigner distributions. For the simulation shown in Fig.~\ref{03048a_free_fit}, the resonant energies of \nuc{11}{Be}*(1.778~MeV) (dot-dashed line) and \nuc{11}{Be}*(2.690~MeV) (dashed line) and their widths (100~keV and 200~keV) were kept constant at the values reported in Ref.~\cite{ajze90} along with the proportional intensities of the two as reported by Pain {\it et al.} in Ref. \cite{pain06prl}. For the third low-decay-energy peak (dotted line), the energy, width, and relative population with respect to the other two resonant level were free parameters. A background distribution due to non-resonant neutrons and neutrons from the direct diffractive breakup channel of \nuc{12}{Be} was included with a Maxwellian distribution, $\sqrt{E} \exp(-E/E_{0})$, where $E_{0}$ was a free parameter (see Ref. \cite{deak87nim} concerning modeling of non-resonant background). The magnitude and $E_{0}$ parameter of the background (double-dot dashed line) were also treated as free parameters, the final best-fit curve with $E_{0}$~= 5.0~MeV was nearly identical to the background curve of Ref. \cite{pain06prl}. The angle and position distributions of the incoming \nuc{12}{Be} beam used in the simulation were adjusted to reproduce the angle and position distributions of the fragments in the charged particle detectors.

\par
Due to a technical failure of the beam counting monitor, it was not possible to extract the cross section directly from the experiment. The overall normalization to extract the cross section of populating the low-energy peak was done by scaling relative to the cross sections reported by Pain {\it et al.}~\cite{pain06prl}. Since the beam energy (39~MeV/nucleon) was much lower than the present experiment (90~MeV/nucleon), the reported cross sections of Pain {\it et al.} were scaled to account for the reduction of knockout cross sections with faster beams. This was done by calculating the single particle cross sections for each state at both energies using the same Eikonal reaction model \cite{tost01npa} that was used in Ref.~\cite{pain06prl}. The single particle cross section ratio for the former to current beam energies is 0.62 for all three states observed: \nuc{11}{Be}*(1.778 \& 2.690 \& 3.949~MeV). The reported cross sections for the 1.778~MeV and 2.690~MeV states were then scaled by this factor of 0.62 (keeping the relative magnitudes constant) and the cross section of the low-energy peak was determined.

\par
The decay energy of the low-energy peak was found to be $S_{n} = 80$(2)~keV as shown in the inset of Fig.~\ref{03048a_free_fit} with a cross section of 15(3)~mb. Systematic uncertainties, due to various beam parameters that fit the measured distributions recorded in the charged-particle detectors, accounts for the limited resolution of fitting the decay width leading to an upper limit of 40~keV that is consistent with the accepted value of 15~keV \cite{ajze90}. The uncertainty of the centroid of the peak is much less affected and a $\chi^2$ analysis yields a 2~keV standard deviation for the uncertainty of the 80~keV value. By adding the measured value of the first excited $2^+$ state in \nuc{10}{Be} at 3.36803(3)~MeV \cite{ajze88} and the recently improved neutron separation energy of 501.3(6)~keV \cite{ring09}, this neutron decay energy corresponds to an excitation energy of 3.949(2)~MeV in \nuc{11}{Be}, and improves the uncertainty of the currently adopted energy of this state (the second $3/2^-$ state at 3.956(15)~MeV in Ref. \cite{ajze90}. The present value is below the value measured by Hirayama {\it et al.} for this state, $3.969^{+0.020}_{-0.009}$~MeV, from \nuc{11}{Li} beta decay \cite{hira05}. The lack of evidence for a resonance below 80~keV shows that the \nuc{11}{Be}*(3.887) state, decaying to the $2^{+}$ in \nuc{10}{Be}, is not measurably populated in the present knockout reaction. 

\par
The large measured cross section of 15(3)~mb for the neutron decay of \nuc{11}{Be}*(3.949) is similar in magnitude to the cross sections for populating \nuc{11}{Be}*(1.778 and 2.69), as reported in Ref.~\cite{pain06prl}. The reported cross sections for populating these two states, after scaling by the single particle cross section ratio (0.62) for the different beam energies, are 19(3) and 14(3)~mb, respectively. The knockout reaction model calculation \cite{tost01npa} yields a single particle cross section of 31.4~mb to populate the second $3/2^-$ state in \nuc{11}{Be}. Haigh {\it et al.} \cite{haig09} measured the decay branching from this 3.949~MeV state to both the ground state (with a decay energy of 3.45~MeV that is outside the geometric acceptance of our setup) and to the $2^{+}$ excited state (the 80~keV channel we measured) of \nuc{10}{Be} with a two-neutron pickup reaction (\nuc{16}{O},\nuc{14}{O}) on \nuc{9}{Be}. Their results show that the branching to these two channels is equal. Earlier work by Hirayama {\it et al.} \cite{hira05} also measured the branching ratio (with large uncertainties) from \nuc{11}{Be}*(3.949) following the beta decay of \nuc{11}{Li}. Therefore, our measured cross section to the first excited state in \nuc{10}{Be} is doubled to get the total single-neutron knockout cross section from  \nuc{12}{Be} to the \nuc{11}{Be}*(3.949) state. This total production cross section of 30(6)~mb leads to a spectroscopic factor of 1.0(2) when compared to the reaction model calculation \cite{tost01npa}. This value is about twice the observed spectroscopic factor of the lower-lying states in \nuc{11}{Be} measured in Refs. \cite{navi00prl,pain06prl}, supporting the interpretation for the character of this $3/2^-$ state as predominantly single-particle, likely due to hole correlations in the $0p_{3/2}$ orbital.

\begin{table*}
\begin{minipage}{\linewidth}
\renewcommand{\footnoterule}{}
\renewcommand{\thefootnote}{\thempfootnote}
\caption{WBP Hamiltonian \cite{wbp} theoretical calculations for the first three $3/2^-$ states in \nuc{11}{Be}. Energies, spectroscopic factors, and their wavefunctions are calculated for the $p$-shell including up to two particles excited into the $sd$-shell \cite{kanu10}.}
\label{spec_calc}
\begin{center}
\begin{minipage}{4.8in}
\begin{tabular*}{\linewidth}%
{|cc|ccc|cc|}%
\hline
E$^*$ Th. & E$^*$ Exp. & & Spec. factors & & Wavefunction & components  \\
(MeV) & (MeV) & from \nuc{12}{Be} g.s. & to \nuc{10}{Be} $0^+$ & to \nuc{10}{Be} $2^+$ & 0$\hbar \omega$ \% & 2$\hbar \omega$ \% \\
 \hline
1.76 & 2.69 & 1.576 & 0.155 & 0.461 & 73 & 27 \\
2.80 & 3.949 & 0.693 & 0.0012 & 0.215 & 19 & 81 \\
4.24 & ? & 0.033 & 0.0053 & 0.221 & 70 & 30 \\
\hline
\end{tabular*}
\end{minipage}
\end{center}
\end{minipage}
\end{table*}

\par
The experimental results can be compared to calculations in the $p$-shell
with the WBP Hamiltonian \cite{wbp} that include 
up to two particles excited into the $sd$-shell \cite{kanu10}.
The wavefunction for the \nuc{12}{Be} $0^{+}$ ground state is calculated to comprise 31\% 0$\hbar \omega$ with $p$-shell configurations and 69\% 2$\hbar \omega$ with two nucleons excited into the into the $sd$-shell. The calculated energies of the first two $3/2^-$ states are about 1~MeV too low compared to their measured values; and experimental energy of a third $3/2^-$ state is not
known, but calculated to be 4.24~MeV. The first $3/2^-$ state in  \nuc{11}{Be} is produced by one nucleon removal from the 0$\hbar \omega$ component of the \nuc{12}{Be} ground state with an observed spectroscopic factor of 0.40(6) \cite{pain06prl} that is significantly smaller than the calculated value of 1.576. The second $3/2^-$ state in  \nuc{11}{Be} (81\% 2$\hbar \omega$) is produced by one-nucleon removal from the 2$\hbar \omega$ component of the  \nuc{12}{Be} ground state. The experimental spectroscopic factor reported herein of 1.0(2) is in reasonable agreement with the calculated value of 0.69. See Table \ref{spec_calc} for more details.

The decay widths are calculated by $\Gamma = C^{2}S \Gamma_{sp}$ where the spectroscopic factors $C^{2}S$ and the single-particle decay widths $\Gamma_{sp}$ are calculated by Eq. 3F-51 in \cite{bohrmott69} using the experimental $Q$ values. The single-particle $l = 1$ decay width for the decay of the first $3/2^{-}$ to the \nuc{10}{Be} $0^+$ ground state ($Q$ = 2.19~MeV) is 1.5~MeV. Combined with the spectroscopic factor of 0.155, the resulting decay width of 0.23~MeV is in good agreement with the experimental value of 0.20(2)~MeV \cite{ajze90}. The single-particle $l = 1$ decay width for the decay of the second $3/2^-$ \nuc{11}{Be}*(3.949) state to the two decay channels, \nuc{10}{Be} $2^+$ ($Q$ = 0.080~MeV) and $0^+$ ($Q$ = 3.448~MeV), are 0.020 and 4.0~MeV, respectively. Combined with the calculated spectroscopic factors; 0.22 for the $2^+$ channel and 0.0012 for the $0^+$ channel, the decay widths are 4.4 and 4.8~keV, respectively. The large variation in spectroscopic factors is due to interference between the various 0$\hbar \omega$ and 2$\hbar \omega$ wavefunction components of the decaying $3/2^-$ state and the $0^+$ or $2^+$ states in \nuc{10}{Be}. The total experimental width is 15(5) keV \cite{ajze90} and, for equal branching ratios \cite{haig09}, the experimental partial widths would each be half that; around 7(3)~keV. The agreement between experiment and theory is surprisingly good, given the small spectroscopic factors involved.

\par
We note that a general feature of analyses of nucleon
knockout reactions is that measured cross sections are
smaller than those calculated using the Eikonal model
with shell-model spectroscopic factors. This empirical
behavior is shown, for example, in Figure 6 of Ref.\
\cite{gade08prc}. The observed reduction factors, $R_s$, show a systematic dependence
on the asymmetry of the neutron and proton separation
energies from the projectile ground state, $\Delta S$.
In the present case, of weakly-bound neutron removal
from $^{12}$Be, the neutron separation energies to the 
$^{11}$Be ground state and 3.949 MeV excited state
correspond to $\Delta S$ of $-$20 MeV and $-$16 MeV,
respectively. These $\Delta S$, and the measured reaction
systematics, suggest $R_s$ values of close to unity in
the present work.


\par
The non-observation of the 3.887~MeV state, decaying preferentially to the $2^{+}$ state in \nuc{10}{Be} by 14~keV, indicates that this state is not strongly populated by single neutron removal from \nuc{12}{Be} or two-neutron transfer \cite{haig09}. This interpretation is also consistent with the results of the three-proton stripping reaction from \nuc{14}{N} \cite{deak87nim} that populated \nuc{11}{Be}*(3.887) but not \nuc{11}{Be}*(3.949), where the likelihood of exciting neutrons to higher sub-shells exists. This 14~keV decay channel was also observed in Ref. \cite{bohl04npa} that selectively populated the 3.887~MeV and 3.949~MeV states by two-proton and two-neutron transfer reactions, respectively. Finally, in another MoNA experiment populating unbound states in \nuc{11}{Be} by the non-selective reaction of direct fragmentation from \nuc{48}{Ca}, neutrons decaying from both excited states near 4~MeV to the $2^{+}$ state in \nuc{10}{Be} were observed \cite{chri08npa}. The similarities between the setups for that experiment and the present supports our interpretation of the selectivity of the single-neutron knockout from \nuc{12}{Be} to \nuc{11}{Be}*(3.949). However, as noted earlier, we cannot rule out the possibility that the \nuc{11}{Be}*(3.887) state is populated and subsequently directly decays predominantly to the ground state of \nuc{10}{Be} by 3.38~MeV neutron decay.


\par
In summary, the resonance observed through neutron-decay spectroscopy measurements of the neutron-unbound excited states in $^{11}$Be at a decay energy of 80(2)~keV indicates the population of the known second  $ 3/2^{-}$ state at 3.949(2)~MeV in \nuc{11}{Be} decaying to the $2^+$ state in \nuc{10}{Be} via neutron emission. The inferred cross section for this decay branch of 15(3)~mb implies a spectroscopic factor near unity for this $ 3/2^{-}$ state, consistent with shell model calculations.

\par
W.A.P. thanks S.~Pain, D. Bardayan, and F.~Nunes for fruitful discussions.
The MoNA project was made possible by
funding from the National Science Foundation
under Grants PHY-0110253, PHY-0132367,
PHY-0132405, PHY-0132434, PHY-0132438,
PHY-0132507, PHY-0132532, PHY-0132567,
PHY-0132641, PHY-0132725, PHY-0758099 , PHY-0098800, and by support
from Ball State University, Central Michigan
University, Concordia College, Florida State
University, Hope College, Indiana University at
South Bend, Michigan State University, Millikin
University, Westmont College, Western
Michigan University, and the National Superconducting Cyclotron
Laboratory. This work was also supported by the NSF grant PHY-06-06007 and by the United Kingdom Science and Technology Facilities Council (STFC) under Grant No. ST/F/012012/1.


\end{document}